%% file: ListUniqueDecodingFoldedSubspaceCodes_arXiv.tex
\documentclass[a4paper, conference]{IEEEtran}

\usepackage[utf8]{inputenc}
\usepackage{amssymb}
\usepackage{amsmath}
\usepackage{color}
\usepackage{theorem}
\usepackage{cite}
\usepackage{url}
\usepackage{hyperref}
\hypersetup{colorlinks=false, pdfborderstyle={/S/U/W 1}}
\usepackage{pgfplots}
\usetikzlibrary{plotmarks}
\usepackage[ruled,vlined,titlenumbered,linesnumbered]{algorithm2e}

\pgfplotsset{
	compat=newest,
  every axis/.append style={
        scale only axis,
	width=0.8\columnwidth,
	height=0.6\columnwidth,
	label style={inner sep=0, font=\normalsize}, 
	tick label style={font=\scriptsize},
	legend style={font=\scriptsize, line width=0.5pt},
	mark size=3,
	mark options={solid, line width=0.8pt},
	major grid style={dashed},
	axis line style = thin}
}

\input{macros.tex}

\newcommand{\KK}{green}
\newcommand{\MV}{black}
\newcommand{\GX}{red}
\newcommand{\FS}{blue}

\IEEEoverridecommandlockouts

\begin{document}

\title{List and Probabilistic Unique Decoding of \\ Folded Subspace Codes}

\author{\IEEEauthorblockN{Hannes Bartz and Vladimir Sidorenko}
\IEEEauthorblockA{Institute for Communications Engineering \\ Technische Universit\"at M\"unchen, Munich, Germany\\
\texttt{hannes.bartz@tum.de, vladimir.sidorenko@tum.de}
}}
\maketitle

\begin{abstract}
 A new class of folded subspace codes for noncoherent network coding is presented.
 The codes can correct insertions and deletions beyond the unique decoding radius for any code rate $R\in\intervallincl{0}{1}$.
 An efficient interpolation-based decoding algorithm for this code construction is given which allows to correct insertions and deletions up to the normalized radius $s\left(1-((1/h+h)/(h-s+1))R\right)$, where $h$ is the folding parameter and $s\leq h$ is a decoding parameter. 
 The algorithm serves as a list decoder or as a probabilistic unique decoder that outputs a unique solution with high probability.
 An upper bound on the average list size of (folded) subspace codes and on the decoding failure probability is derived.
 A major benefit of the decoding scheme is that it enables probabilistic unique decoding up to the list decoding radius.
\end{abstract}


\section{Introduction}\label{sec:introduction}
Subspace codes have been proposed for error control for noncoherent random linear network coding, e.g. when the network topology and the in-network linear combinations are not known by the transmitter and the receiver \cite{koetter_kschischang, silva_rank_metric_approach}.
K\"otter and Kschischang proposed a Reed--Solomon like construction based on rank-metric codes (referred to as KK codes) that can be decoded efficiently \cite{koetter_kschischang}.
List decodable variants of subspace codes have been proposed in \cite{Mahdavifar2010Algebraic, GuruswamiInsertionsDeletions, GuruswamiWang13Explicit, BartzWachterInterleavedSubspace2014, TrautmannSilbersteinRosenthal-ListDecodingLiftedGabidulinCodes, WachterzehZeh-InterpolationInterleavedGabidulin, GuruswamiXing-ListDecodingRSAGGabidulinSubcodes_2012} and allow to correct insertions and deletions beyond half the minimum subspace distance.
The challenge of list decoding subspace codes is to decrease the size of the list of candidate codewords, which is exponential in the dimension of the transmitted subspace \cite{Wachterzeh_BoundsListDecodingRankMetric_IEEE-IT_2013}.
Most list decodable subspace codes are based on KK codes and control the size of the list by restricting the message symbols or the code locators to belong to a subfield.
Guruswami and Wang \cite{GuruswamiXing-ListDecodingRSAGGabidulinSubcodes_2012} showed that punctured subspace codes can be list decoded up to the theoretical limit for any code rate.
The list size for this decoder is further reduced in \cite{GuruswamiWang13Explicit} by applying hierarchical subspace evasive sets.
The output of this decoder is a \emph{basis} for the affine space of candidate solutions resulting in a very large list of exponential size in the dimension of the transmitted subspace with high probability.

In this paper we define a new class of folded subspace codes that can be decoded from insertions and deletions for any code rate.
We present an interpolation-based decoding algorithm that can be used as a list decoder or as a probabilistic unique decoder.
Both schemes can correct insertions and deletions beyond half the minimum subspace distance for any code rate. 
The probabilistic unique decoder returns a unique solution with high probability and requires at most $\OCompl{s^2\nReceive^2}$ operations in $\Fqm$, where $s$ is a decoding parameter (small integer) and $\nReceive$ is the dimension of the received subspace. 
The decoding scheme is well suited for practical application. 
We give an upper bound on the probability of a decoding failure (i.e. a list of size larger than one) and verify the results by simulations.

This paper is structured as follows. 
In Section~\ref{sec:preliminaries}, we describe the notation and give basic definitions.
Section~\ref{sec:principleFS} introduces a new class of folded subspace codes and presents an efficient interpolation-based decoding scheme.
In Section~\ref{sec:listuniqueFS} we apply the algorithm to list and unique decoding of folded subspace codes and highlight the improvements of the decoding scheme.
Finally, Section~\ref{sec:conclusion} concludes this paper.

\section{Preliminaries}\label{sec:preliminaries}
\subsection{Finite Fields and Subspaces}
Let $q$ be a power of a prime, and let $\Fq$ be the finite field of order $q$ and let $\Fqm$ be its extension field of degree $m$. 
Any element from $\Fqm$ can be represented by a row vector of length $m$ over $\Fq$ for a fixed basis.
By $\Fq^N$ we denote a vector space of dimension $N$ over $\Fq$ and the set of all subspaces of $\Fq^N$ is the projective space $\ProjspaceAny{N}$.
The set of all $\ell$-dimensional subspaces of $\Fq^N$ is the \emph{Grassmannian} and is denoted by $\Grassm{N,\ell}$.
We denote matrices and vectors by bold uppercase and lowercase letters such as $\vec{A}$ and $\Mat{a}$ and index their elements beginning from zero. 
The rank of a matrix $\Mat{A} \in \Fq^{m \times n}$ is denoted by $\rk(\Mat{A})$ and the kernel of $\Mat{A}$ is denoted by $\ker(\Mat{A})$.
The row space of a set of vectors $\set{B}$ over $\Fq$ is denoted by $\Rowspace{\set{B}}$.
For two subspaces $\myspace{U},\myspace{V}\in\ProjspaceAny{N}$, the direct sum $\myspace{U}\oplus\myspace{V}$ is the smallest subspace containing both $\myspace{U}$ and $\myspace{V}$. 
The \emph{subspace distance} between $\myspace{U},\myspace{V}$ in $\ProjspaceAny{N}$ is
\begin{align}\label{eq:subspaceDistance}
\setlength{\abovedisplayskip}{3pt}
\setlength{\belowdisplayskip}{3pt}
	\Subspacedist{\myspace{U},\myspace{V}}
	  &=\dim(\myspace{U})+\dim(\myspace{V})-2\dim(\myspace{U}\cap \myspace{V}). 
\end{align}
A \emph{subspace code} is a nonempty subset of $\ProjspaceAny{N}$, and has minimum subspace distance $d_s$ when all subspaces in the code have distance larger than or equal to $d_s$ from each other. 

As channel model we use the operator channel from \cite{koetter_kschischang}.
Such a channel has input and output alphabet $\ProjspaceAny{N}$. 
The output $\myspace{U}$ is related to the input $\myspace{V}$ with $\dim(\myspace{V})=\nTransmit$ by
\begin{equation}
\setlength{\abovedisplayskip}{3pt}
\setlength{\belowdisplayskip}{3pt}
 \myspace{U}=\mathcal{H}_{\nTransmit-\deletions}(\myspace{V})\oplus \myspace{E}
\end{equation}
where $\mathcal{H}_{\nTransmit-\deletions}(\myspace{V})$ returns a random $(\nTransmit\!-\!\deletions)$-dimensional subspace of $\myspace{V}$, and $\myspace{E}$ denotes an error space of dimension~$\insertions$ with $\myspace{V}\cap\myspace{E}=\emptyset$.
The distribution of $\mathcal{H}_{\nTransmit-\deletions}(\myspace{V})$ does not affect the performance of the code and can be chosen to be uniform (see \cite{koetter_kschischang}).
The dimension of the received subspace $\myspace{U}$ is thus $\nReceive\!=\!\nTransmit\!-\!\deletions\!+\!\insertions$ and we call $\deletions$ the number of \emph{deletions} and $\insertions$ the number of \emph{insertions}.

\subsection{Linearized Polynomials}
For any element $a\in\Fqm$ and any integer $i$ let $a^{[i]}\defeq a^{q^{i}}$ be the Frobenius power of $a$.
A nonzero polynomial of the form
$p(x)=\sum_{i=0}^{d}p_ix^{[i]}$
with $p_i\in \Fqm$, $p_d\neq 0$, is called a \emph{linearized polynomial} of $q$-degree $\deg_q(p(x))=d$, see \cite{Ore_OnASpecialClassOfPolynomials_1933,Lidl-Niederreiter:FF1996}.
Evaluating a linearized polynomial forms a linear map over $\Fq$, i.e. for all $a,b\in\Fq$ and $x_1,x_2\in\Fqm$, we have $p(ax_1+bx_2)=ap(x_1)+bp(x_2)$. 
The noncommutative composition $f(x)\otimes g(x)=f(g(x))$ of two linearized polynomials $f(x)$ and $g(x)$ of $q$-degree $d_1$ and $d_2$ is a linearized polynomial of $q$-degree $d_1+d_2$.
The set of all linearized polynomials over $\Fqm$ forms a noncommutative ring $\Linpolyring$ with identity under addition ``+'' and composition ``$\otimes$''. 
The \qVan matrix of the vector $\vec{a}=\vecelements{a}\in \Fqm^n$ is defined as
\begin{equation}
 \setlength{\abovedisplayskip}{8pt}
 \setlength{\belowdisplayskip}{8pt}
  \Mooremat{r}{\vec{a}}=\MoormatExplicit{a}{r}{n}.
\end{equation}
The rank of $\Mooremat{r}{\vec{a}}$ is $\min\{r,n\}$ if the elements $a_0,\dots,a_{n-1}$ are linearly independent over \Fq, see \cite{Lidl-Niederreiter:FF1996}.

\section{Interpolation-Based Decoding of Folded Subspace Codes}\label{sec:principleFS}
We present a new construction of folded subspace (FS) codes that can be decoded from insertions and deletions beyond the unique decoding radius for any code rate $R$. 
This work is motivated by the constructions in \cite{Mahdavifar2012Listdecoding} and \cite{GuruswamiWang2013LinearAlgebraic}.

Let $\pe$ be a primitive element of the field $\Fqm$ with polynomial basis $\pe^0,\pe^1,\dots,\pe^{m-1}$ over $\Fq$. 
\begin{definition}\label{def:foldedSubspace}
 An $h$-folded subspace code \FSub{h;\nTransmit,k} of dimension $\nTransmit$, where $h\nTransmit\leq m$, is defined as the set of subspaces 
 \begin{equation*}
 \setlength{\abovedisplayskip}{10pt}
 \setlength{\belowdisplayskip}{10pt}
	\Rowspace{\left\{ \big(\pe^{jh}, f(\pe^{jh}), f(\pe^{jh+1}), \dots, f(\pe^{(j+1)h-1})\big):j\in\set{J}\right\}}
 \end{equation*}
 for all $f(x)\in\Linpolyring$, $\deg_q(f(x))<k$, where $\set{J}= \intervallexcl{0}{\nTransmit}$.
\end{definition}
The dimension of the ambient space 
 \begin{equation*}
 \setlength{\abovedisplayskip}{6pt}
 \setlength{\belowdisplayskip}{5pt}
	 W_s\!= \Rowspace{\pe^0,\pe^h,\dots,\pe^{(\nTransmit-1)h}} \oplus \underbrace{\Fqm\oplus \dots \oplus \Fqm}_{h\:\text{times}}
 \end{equation*}
 is $N=\nTransmit+hm$, since the vectors in the space $\Rowspace{\pe^0,\pe^h,\dots,\pe^{(\nTransmit-1)h}}$ have nonzero components at the $\nTransmit$ known positions $0,h,2h,\dots,(\nTransmit-1)h$ only. 
 The zeroes at the known positions do not need to be transmitted and can be inserted at the receiver. The code rate is $R=\frac{km}{\nTransmit(\nTransmit+hm)}$.
\begin{lemma}\label{lem:subDistFS}
The minimum subspace distance of the code \FSub{h;\nTransmit,k} is $d_{S,min}=2(\nTransmit-\lceil\frac{k}{h}\rceil+1)$.
\end{lemma}

\begin{IEEEproof}
 Let $\myspace{V}$ and $\myspace{V'}$ be two distinct codewords generated by $f(x)$ and $g(x)$ with $q$-degrees less than $k$ and suppose $\dim(\myspace{V}\cap\myspace{V'})\geq\lceil\frac{k}{h}\rceil$.
 Then $f(x)$ and $g(x)$ must agree on $h\lceil\frac{k}{h}\rceil\geq k$ linearly independent points, which is not possible since the $q$-degree of both polynomials is less than $k$.
 Thus the dimension of the intersection space $\myspace{V}\cap\myspace{V'}$ can be at most $\lceil\frac{k}{h}\rceil-1$.
 Using \eqref{eq:subspaceDistance} we have
 \begin{align*}
  \Subspacedist{\myspace{V},\myspace{V'}}=2\nTransmit-2\dim(\myspace{V}\cap \myspace{V'}) =2(\nTransmit-\Big\lceil\frac{k}{h}\Big\rceil-1).
 \end{align*}
\end{IEEEproof}

\subsection{Interpolation Step}\label{subsec:interpolationFS}
Suppose we receive a basis of dimension $\nReceive = \nTransmit - \deletions + \insertions$  
\begin{equation*}
  \setlength{\abovedisplayskip}{4pt}
  \setlength{\belowdisplayskip}{4pt}
  \{(x_j, y_{j,0}, y_{j,1}, \dots, y_{j,h-1}) : j\in\intervallexcl{0}{\nReceive}\}
\end{equation*}
of the received subspace $\myspace{U}$.
Let the matrix $[\vec{x}^T, \vec{y}^{(1)T}, \dots, \vec{y}^{(h)T}] \in \Fqm^{\nReceive \times (h+1)}$ contain this basis as rows.
Suppose we receive a $\Fq$-linear combination of the transmitted basis vectors of the form
\begin{equation*}
 \sum_{j=0}^{\nTransmit-1}\!\!\lambda_j \left(\pe^{jh}, f\big(\pe^{jh}\big), f\big(\pe^{jh+1}\big),\dots, f\big(\pe^{(j+1)h-1}\big)\right)
\end{equation*}
with $\lambda_j\in\Fq$.
Due to the linear property of linearized polynomials we can rewrite this as
\begin{align}\label{eqReceivedBasis}
	\Bigg(\sum_{j=0}^{\nTransmit-1}\!\!\lambda_j\pe^{jh}, f\big(\sum_{j=0}^{\nTransmit-1}\!\!\lambda_j\pe^{jh}\big), 
	      \dots, f\big(\pe^{h-1}\sum_{j=0}^{\nTransmit-1}\!\!\lambda_j\pe^{jh}\big)\Bigg).
\end{align}

For the interpolation step we must solve the following problem.

\begin{problem}\label{prob:intProblemFS}
Given the integers $D$ and $s\leq h, 1\leq s\leq h$, find a nonzero $(s+1)$-variate linearized polynomial of the form 
	\begin{equation}\label{eq:intPolyFS}
	\intPoly = Q_0(x) + Q_1(y_1) + \dots + Q_s(y_s),
	\end{equation}
which satisfies for all $i \in \intervallincl{0}{h-s}, j \in \intervallincl{0}{\nReceive-1}$:
	\begin{itemize}
	\item[$\bullet$] $Q(x_j\pe^{i}, y_{j,i}, y_{j,i+1},\dots,y_{j, i+s-1}) = 0$,
	\item[$\bullet$] $\deg_q(Q_0(x)) < \degConstraint$,
	\item[$\bullet$] $\deg_q(Q_\ell(y_\ell))< \degConstraint-(k-1)$, $\forall \ell \in \intervallincl{1}{s}$.
	\end{itemize}
\end{problem}
Here we use (\ref{eqReceivedBasis}) to determine the code locators for the $(h-s+1)$ interpolation tuples for each dimension as $x_j\pe^{i}, \forall i \in \intervallincl{0}{h-s},j \in \intervallincl{0}{\nReceive-1}$.  
A solution to Problem~\ref{prob:intProblemFS} can be found by solving a homogeneous linear system of equations.
Denote the polynomials of~\eqref{eq:intPolyFS} by
$Q_0(x)=\sum_{j=0}^{\degConstraint-1}q_{0,j}x^{[j]}$ and 
$Q_i(y_i)=\sum_{j=0}^{\degConstraint-k}q_{i,j}y_i^{[j]}$.
Let the matrix $\Mat{T}$ contain all $\nReceive(h-s+1)$ interpolation tuples $\left(x_j\pe^{i}, y_{j,i}, y_{j,i+1},\dots,y_{j, i+s-1}\right)$, $\forall i \in \intervallincl{0}{h-s}, j \in \intervallincl{0}{\nReceive-1}$ as rows and denote by $\vec{t}_\ell$ the $\ell$-th column of $\Mat{T}$ for $\ell\in\intervallincl{0}{s}$.
The coefficients $q_{i,j}$ can be found by solving a linear system 
\begin{equation}\label{eq:intSystemFS}
   \setlength{\abovedisplayskip}{4pt}
   \setlength{\belowdisplayskip}{4pt}
 \Mat{R}\cdot\vec{q}_I^T=\Mat{0} 
\end{equation}
where $\Mat{R}$ is an $\nReceive(h-s+1)\times \degConstraint(s+1)-s(k-1)$ matrix:
\begin{align}\label{eq:intMatrixFS}
 \Mat{R}\!=\!\Big(\Mooremat{\degConstraint}{\vec{t}_{0}^T}^T,\Mooremat{\degConstraint-k+1}{\vec{t}_{1}^T}^T,\dots,\Mooremat{\degConstraint-k+1}{\vec{t}_{s}^T}^T\Big)
\end{align}
and 
$\vec{q}_I=(q_{0,0},\dots,q_{0,\degConstraint-1}| \ 
		\dots \ |q_{s,0},\dots,q_{s,\degConstraint-k})$.
\begin{lemma}
 A nonzero polynomial fulfilling the interpolation constraints in Problem \ref{prob:intProblemFS} exists if
 \begin{equation}\label{degreeConstraint}
   \setlength{\abovedisplayskip}{3pt}
   \setlength{\belowdisplayskip}{3pt} 
   \degConstraint=\Bigg\lceil\frac{\nReceive (h-s+1)+s(k-1)+1}{s+1}\Bigg\rceil.
 \end{equation}
\end{lemma}

\begin{IEEEproof}
 Problem \ref{prob:intProblemFS} forms a homogeneous linear system of $\nReceive(h\!-\!s\!+\!1)$ equations in $\degConstraint(s\!+\!1)\!-\!s(k\!-\!1)$ unknowns.
 This system has a nonzero solution if the number of linear independent equations is less than the number of unknowns, i.e., if
 \vspace*{-7pt}
 \begin{align}\label{eq:degreeConditionFS} 
	\nReceive (h-s+1)&<\degConstraint(s+1)-s(k-1)	\\
			\Longleftrightarrow\qquad \degConstraint&\geq\frac{\nReceive (h-s+1)+s(k-1)+1}{s+1}	\nonumber. \\[-25pt]\nonumber	
 \end{align}
\end{IEEEproof}
The receiver knows $\nReceive$, the code parameter $k$ and the decoding parameter $s$ and can compute the degree restriction $\degConstraint$ in \eqref{degreeConstraint}.

\begin{theorem}
 Let $\intPoly\neq 0$ fulfill the interpolation constraints in Problem \ref{prob:intProblemFS}. 
 If 
 \begin{equation}\label{eq:decRadiusFS}
   \setlength{\abovedisplayskip}{4pt}
   \setlength{\belowdisplayskip}{2pt}
  \insertions + s\deletions < s\left(\nTransmit - \frac{k-1}{h-s+1}\right)
 \end{equation}
 then
 \begin{equation}\label{eq:rootFindingPolyFS}
     \setlength{\abovedisplayskip}{3pt}
   \setlength{\belowdisplayskip}{3pt}
  P(x)\defeq Q(x,f(x),f(\pe x),\dots, f(\pe^{s-1} x))=0.
 \end{equation}
\end{theorem} 

\begin{IEEEproof}
 The dimension of the noncorrupted subspace $\myspace{V}\cap\myspace{U}$ is $\nTransmit-\deletions$.
 The code locators of the noncorrupted dimensions are linearly independent and thus we have $(\nTransmit-\deletions)(h-s+1)$ linearly independent interpolation points (roots) in \eqref{eq:intPolyFS}.
 Since $\deg_q(P(x))<\degConstraint$ the dimension of the root space of $P(x)$ is at most $\degConstraint-1$.
 If
 \begin{equation}\label{eq:decConditionFS}
   \setlength{\abovedisplayskip}{4pt}
   \setlength{\belowdisplayskip}{4pt}
	\degConstraint\leq(\nTransmit-\deletions)(h-s+1)
 \end{equation}
 then $P(x)$ has more linearly independent roots than its degree.
 This is possible only if $P(x)=0$.
 Combining \eqref{eq:degreeConditionFS} and \eqref{eq:decConditionFS} and using
 \begin{equation}
  \setlength{\abovedisplayskip}{2pt}
  \setlength{\belowdisplayskip}{2pt}
  \nReceive=\nTransmit+\insertions-\deletions
 \end{equation}
 we get \eqref{eq:decRadiusFS}.
\end{IEEEproof}
For $h=s=1$ (no folding) we have $\insertions + \deletions < \nTransmit-k+1$
which is identical to the decoding radius of KK codes in \cite{koetter_kschischang}.
For $s=h$ the algorithm in \cite{Mahdavifar2012Listdecoding} is identical to the proposed scheme.
By using the approximation $R\approx\frac{(k-1)m}{\nTransmit(\nTransmit+hm)}$ the normalized decoding radius $\tau_f=\frac{\insertions+s\deletions}{\nTransmit}$ is given by
\begin{align*}
 \tau_f&\approx s\left(1-\frac{\nTransmit+hm}{m(h-s+1)}R\right).
\end{align*}
If $\nTransmit h\approx m$ then we may write 
$\tau_f\approx s\left(1-\frac{1/h+h}{h-s+1}R\right)$.

Problem~\ref{prob:intProblemFS} can be solved by the efficient interpolation algorithm in \cite{Xie2011General} requiring at most $\OCompl{s^2\nReceive \degConstraint(h-s+1)}<\OCompl{s^2\nReceive^2}$ operations in $\Fqm$.

\subsection{Root-Finding Step}\label{subsec:root-findingFS}
Given a polynomial $\intPoly$, we must find all polynomials $f(x)\in\Linpolyring$ of degree less than $k$ which are a solution to \eqref{eq:rootFindingPolyFS}.
To increase the probability to find a unique solution we use a similar idea as in \cite{WachterzehZeh-InterpolationInterleavedGabidulin, BartzWachterInterleavedSubspace2014}. 
The solution space of the interpolation system \eqref{eq:intSystemFS} has dimension larger than one in general.
In this case, there exists a set of linearly independent linearized polynomials $\intPoly$ which are a solution to Problem~\ref{prob:intProblemFS}. 
Instead of one polynomial we use a basis for the solution space of \eqref{eq:intSystemFS} to increase the probability that the root-finding system has a unique solution.
We now derive a lower bound on the dimension of the solution space of \eqref{eq:intSystemFS}.
\begin{lemma}\label{lem:rankIntMatrixFS}
The dimension $d_I$ of the solution space of the interpolation system \eqref{eq:intSystemFS} satisfies $d_I\geq s(\degConstraint-k+1)-\insertions(h-s+1)$.
\end{lemma}

\begin{IEEEproof}
Let $\Mat{T}'$ contain the $(\nTransmit-\deletions)(h-s+1)$ noncorrupted interpolation tuples as rows and denote by $\vec{t}_\ell^\prime$ the $\ell$-th column of $\Mat{T}^\prime$ for $\ell\in\intervallincl{0}{s}$.
Assume w.l.o.g. that the first $(\nTransmit-\deletions)(h-s+1)$ rows of $\Mat{R}$ correspond to the noncorrupted interpolation tuples and denote this matrix by $\Mat{R}^\prime$.
The first $\degConstraint$ columns of $\Mat{R}^\prime$ form a $(\nTransmit-\deletions)(h-s+1)\times \degConstraint$ \qVan matrix $\Mooremat{\degConstraint}{\vec{t}_0^{\prime\; T}}^T$ of rank $\degConstraint$ since the elements in $\vec{t}_0^\prime$ are linearly independent and \eqref{eq:decConditionFS} holds.
The $s$ \qVan matrices $\Mooremat{\degConstraint-(k-1)}{\vec{t}_1^{\prime\; T}}^T,\dots, \Mooremat{\degConstraint-(k-1)}{\vec{t}_s^{\prime\; T}}^T$ are linear combinations of the rows of $\Mooremat{\degConstraint}{\vec{t}_0^{\prime\; T}}^T$ and hence do not increase the rank.
$(h-s+1)$ interpolation constraints (i.e. rows) are added to $\Mat{R}'$ for every malicious dimension.
Thus $\insertions$ insertions can increase the rank of $\Mat{R}'$ by at most $\insertions (h-s+1)$.
Hence we have $\rk(\Mat{R})\leq \degConstraint+\insertions (h-s+1)$.
The dimension of the solution space of the interpolation system $d_I\!:=\!\dim \ker(\Mat{R})$ is 
\begin{equation*}
	d_I\geq \degConstraint(s\!+\!1)\!-\!s(k\!-\!1)\!-\!\rk(\Mat{R})=s(\degConstraint\!-\!k\!+\!1)\!-\!\insertions(h\!-\!s\!+\!1).
	\vspace*{-5pt}	
\end{equation*}
\end{IEEEproof}

We now set up the root-finding system using $d_I$ polynomials $\intPoly$.
Define the polynomials 
\begin{equation*}
 B_i^{(\ell)}(x)=q_{1,i}^{(\ell)}+q_{2,i}^{(\ell)}x+q_{3,i}^{(\ell)}x^2+\dots+q_{s,i}^{(\ell)}x^{(s-1)}
\end{equation*}
for $\ell\!\in\!\intervallincl{1}{d_I}$ and the vectors $\vec{b}_{i,j}\!=\!\left(B_i^{(1)}(\pe^{[j]}) \dots B_i^{(d_I)}(\pe^{[j]})\right)^T$
and $\vec{q}_{0,i}=\left(q_{0,i}^{(1)} \ \dots \ q_{0,i}^{(d_I)}\right)$
for $i,j\in\intervallincl{0}{k-1}$.
The root-finding matrix is
\begin{align}\label{eq:rootFindingFS}
\Mat{B}=
 \begin{pmatrix}
  \vec{b}_{0,0}			&				&	 &	\\
  \vec{b}_{1,1}^{[-1]}		& \vec{b}_{0,1}^{[-1]}		&	 &	\\[-4pt]
  \vdots			&	\dots			& \ddots &	\\
  \vec{b}_{k-1,k-1}^{[-(k-1)]}	& \vec{b}_{k-2,k-1}^{[-(k-1)]}	& \dots  &\vec{b}_{0,k-1}^{[-(k-1)]}
 \end{pmatrix}
\end{align}
and $\vec{q}=\left(\vec{q}_{0,0} \ \vec{q}_{0,1}^{[-1]} \ \dots \ \vec{q}_{0,k-1}^{[-(k-1)]}\right)^T$.

We can find the coefficients of the message polynomial $f(x)$ by solving the linear system
\begin{equation}\label{eq:RFsystemFS}
 \Mat{B}\cdot \vec{f}=-\vec{q}
 \vspace*{-3pt}
\end{equation}
where $f(x)$ is connected with the vector $\vec{f}$ by $\vec{f}=\left(f_0 \ f_1^{[-1]} \ \dots \ f_{k-1}^{[-(k-1)]}\right)^T\!$.
The root-finding system \eqref{eq:RFsystemFS} has at least one solution, i.e. $\vec{q}$ is always in the column space of $\Mat{B}$ since we guarantee that the transmitted message polynomial $f(x)$ is a solution to \eqref{eq:rootFindingPolyFS} if $\insertions$ and $\deletions$ satisfy \eqref{eq:decRadiusFS}.
Due to the lower triangular structure of $\Mat{B}$ the root-finding system \eqref{eq:RFsystemFS} can be solved in at most $\OCompl{k^2}$ operations in $\Fqm$.

\section{List and Unique Decoding of Folded Subspace Codes}\label{sec:listuniqueFS}
We now show how the interpolation-based decoding scheme from Section~\ref{sec:principleFS} can be used as a list decoder and as a probabilistic unique decoder.
We focus on the probabilistic unique decoding approach that is well suited for applications.

\subsection{List Decoding Approach}
The solution space of the root-finding system \eqref{eq:RFsystemFS} is an affine subspace over $\Fq$.
In case $\Mat{B}$ in \eqref{eq:RFsystemFS} has rank less than $k$, we obtain a list of possible message polynomials $f(x)$ that satisfy \eqref{eq:rootFindingPolyFS}.
\begin{lemma}\label{lem:maxListSize}
 The dimension of the affine solution space of \eqref{eq:RFsystemFS} is at most $q^{m(s-1)}$.
\end{lemma}

\begin{IEEEproof}
 The lower triangular root-finding matrix $\Mat{B}$ has full rank if and only if all diagonal elements $\vec{b}_{0,0},\dots, \vec{b}_{0,k-1}$ are nonzero vectors.
 The entries of each $\vec{b}_{0,i}$ are the evaluations of $d_I$ polynomials of degree at most $s-1$ at $\pe^{[i]},i\in\intervallincl{0}{k-1}$.
 Since the conjugates $\pe, \pe^{[1]},\dots,\pe^{[k-1]}$ are all distinct and $\deg(B_0^{(\ell)}(x))<s$ for all $\ell\in\intervallincl{1}{d_I}$, we can have $\vec{b}_{0,i}=\vec{0},i\in\intervallincl{0}{k-1}$ at most $(s-1)$ times. 
 For each $\vec{b}_{0,i}=\vec{0}, i\in\intervallincl{0}{k-1}$ the coefficient $f_i$ can be any element in $\Fqm$.
 Thus the dimension of the affine solution space is at most $q^{m(s-1)}$.
\end{IEEEproof}

Using $d_I$ polynomials for the root-finding step does not reduce the worst case list size.
The probability that $\Mat{B}$ is nonsingular increases with $d_I$ and thus the \emph{average} list size is reduced. 

\begin{lemma}\label{lem:volSubBall}
  The number $\volSubBall{\nReceive,\nTransmit,\tau}$ of $\nTransmit$-dimensional subspaces in $\ProjspaceAny{N}$ at subspace distance at most $\tau$ from a fixed $\nReceive$-dimensional subspace in $\ProjspaceAny{N}$ is
  \begin{equation}\label{eq:volSubBallTau}
   \setlength{\abovedisplayskip}{3pt}
   \setlength{\belowdisplayskip}{3pt}
   \volSubBall{\nReceive,\nTransmit,\tau}=\sum_{j=u_l}^{u_m}q^{j(j-\nReceive+\nTransmit)}\quadbinom{\nReceive}{j}\quadbinom{N-\nReceive}{j-\nReceive+\nTransmit}
  \end{equation}
  where $u_l=\lceil\frac{\nReceive-\nTransmit}{2}\rceil$ and $u_m=\lfloor\frac{\nReceive-\nTransmit+\tau}{2}\rfloor$.
\end{lemma}

\begin{IEEEproof}
 Denote by $\numSubShell{\nReceive,\nTransmit,t}$ the number of $\nTransmit$-dimensional subspaces in subspace distance \emph{exactly} $t$ from an $\nReceive$-dimensional subspace.
 In \cite[Lemma~2]{Gadouleau2010SubspacePacking} it is shown that 
 \begin{equation}\label{eq:numSubD}
   \setlength{\abovedisplayskip}{3pt}
  \setlength{\belowdisplayskip}{3pt}
  \numSubShell{\nReceive,\nTransmit,t}=q^{u(t-u)}\quadbinom{\nReceive}{u}\quadbinom{N-\nReceive}{t-u} 
 \end{equation}
 if $u=\frac{\nReceive-\nTransmit+t}{2}$ is an integer and $0$ otherwise.
 The number of all $\nTransmit$-dimensional subspaces at distance \emph{at most} $\tau$ is
 \begin{align}\label{eq:volSubBall}
   \setlength{\abovedisplayskip}{3pt}
  \setlength{\belowdisplayskip}{3pt}
   \volSubBall{\nReceive,\nTransmit,\tau}&=\sum_{t=0}^{\tau}\numSubShell{\nReceive,\nTransmit,t}.	\nonumber 
 \end{align}
 Since $\numSubShell{\nReceive,\nTransmit,t}\neq 0$ if and only if $u=\frac{\nReceive-\nTransmit+t}{2}$ is an integer we rewrite \eqref{eq:numSubD} in terms of $u$. 
 Substituting $t=2u-\nReceive+\nTransmit$ in \eqref{eq:numSubD} we obtain the limits $u_l=\lceil\frac{\nReceive-\nTransmit}{2}\rceil$ and $u_m=\lfloor\frac{\nReceive-\nTransmit+\tau}{2}\rfloor$ and get \eqref{eq:volSubBallTau}.
\end{IEEEproof}

\begin{theorem}\label{thm:avgListSizeFS}
 Let \FSub{h;\nTransmit,k} be a constant dimension subspace code over $\Fqm$ and let $N=\nTransmit+hm$ be the dimension of the ambient vector space.
 Let the number of insertions $\insertions$ and deletions $\deletions$ fulfill \eqref{eq:decRadiusFS}.
 The average list size $\bar{L}_f(\tau)$, i.e. the average number of codewords at subspace distance at most $\tau=\insertions+s\deletions$ from a received $\nReceive$-dimensional subspace satisfies 
 \begin{equation*}
   \setlength{\abovedisplayskip}{8pt}
   \setlength{\belowdisplayskip}{1pt}
   \bar{L}(\tau)<1+16(\frac{\tau}{2}\!+\!1)q^{mk+(\nReceive-\lfloor\frac{\nReceive-\nTransmit+\tau}{2}\rfloor)(\nTransmit+\lfloor\frac{\nReceive-+\tau}{2}\rfloor-N)}.
 \end{equation*}
\end{theorem}

\begin{IEEEproof}
 Let the received subspace $\myspace{Y}$ be chosen uniformly at random from all subspaces in the Grassmannian $\Grassm{N,\nReceive}$.
 The number of $\nTransmit$-dimensional subspaces in subspace distance at most $\tau$ from $\myspace{Y}$ is $\volSubBall{\nReceive,\nTransmit,\tau}$.
 If $\tau$ satisfies \eqref{eq:decRadiusFS} we know that the causal (transmitted) codeword is in subspace distance at most $\tau$ from $\myspace{Y}$. 
 There are $q^{mk}-1$ noncausal codewords (subspaces) out of $\quadbinom{N}{\nTransmit}$ possible $\nTransmit$-dimensional subspaces.

 Thus there are on average 
 \begin{align*}
  \setlength{\abovedisplayskip}{3pt}
  \setlength{\belowdisplayskip}{3pt}
  \bar{L}'(\tau)=(q^{mk}-1)\frac{\volSubBall{\nReceive,\nTransmit,\tau}}{\quadbinom{N}{\nTransmit}}
 \end{align*} 
 noncausal codewords in subspace distance at most $\tau$ from the received subspace.
 Let $u_l=\lceil\frac{\nReceive-\nTransmit}{2}\rceil$ and $u_m=\lfloor\frac{\nReceive-\nTransmit+\tau}{2}\rfloor$.
 Using Lemma~\ref{lem:volSubBall} and the approximation $q^{\ell(n-\ell)}<\quadbinom{n}{\ell}<4q^{\ell(n-\ell)}$ (see \cite{koetter_kschischang}) we have 
 \begin{align*}
 \bar{L}'(&\tau)\!=(q^{mk}-1)\frac{\volSubBall{\nReceive,\nTransmit,\tau}}{\quadbinom{N}{\nTransmit}}
 \\
 <&\frac{q^{mk}}{q^{\nTransmit(N-\nTransmit)}}\sum_{j=u_l}^{u_m}q^{j(j-\nReceive+\nTransmit)}\quadbinom{\nReceive}{j}\quadbinom{N-\nReceive}{j-\nReceive+\nTransmit}	
 \\
 <&q^{mk-\nTransmit(N-\nTransmit)}\cdot\underbrace{(u_m\!-\!u_l\!+\!1)}_{<\frac{\tau}{2}+1}\cdot q^{u_m(u_m-\nReceive+\nTransmit)}
 \\
 &\hspace*{120pt}\cdot\quadbinom{\nReceive}{u_m}\quadbinom{N-\nReceive}{u_m-\nReceive+\nTransmit} 
 \\
 <&q^{mk-\nTransmit(N-\nTransmit)}\cdot \left(\frac{\tau}{2}\!+\!1\right)\cdot q^{u_m(u_m-\nReceive+\nTransmit)}
 \\
 &\cdot 4q^{u_m(\nReceive-u_m)} \cdot 4q^{(u_m-\nReceive+\nTransmit)(N-\nReceive-(u_m-\nReceive+\nTransmit))} 
 \\
 =&16\left(\frac{\tau}{2}\!+\!1\right)q^{mk-\nTransmit(N-\nTransmit)+u_m\nTransmit + (u_m-\nReceive+\nTransmit)(N-u_m-\nTransmit)} 
 \\
 =&16\left(\frac{\tau}{2}\!+\!1\right)q^{mk+(\nReceive-u_m)(\nTransmit+u_m-N)} 
 \\
 =&16\left(\frac{\tau}{2}\!+\!1\right)q^{mk+(\nReceive-\lfloor\frac{\nReceive-\nTransmit+\tau}{2}\rfloor)(\nTransmit+\lfloor\frac{\nReceive-+\tau}{2}\rfloor-N)} 
\end{align*}
 Including the causal codeword we get $\bar{L}(\tau)\!=\!1\!+\!\bar{L}'(\tau)$. 
\end{IEEEproof}

\subsection{Probabilistic Unique Decoder}
In the worst case the decoder outputs an exponential number of candidate message polynomials.
We show that a list of size larger than one is a rare event. 
This allows us to use the algorithm as a probabilistic unique decoder which returns a unique solution or a decoding failure in case the list size is larger than one, i.e. $\rk(\Mat{B})<k$.

The root-finding system \eqref{eq:RFsystemFS} has a unique solution if the rank of $\Mat{B}$ is full. 
This is fulfilled if and only if at least one entry of each $\vec{b}_{0,i},i\in\intervallincl{0}{k-1}$ is nonzero.

\begin{lemma}\label{lem:failProb}
Denote by $d_I$ the dimension of the solution space of \eqref{eq:intSystemFS}.
Then the decoding failure probability is upper bounded by
\begin{equation}\label{eq:fracNonCorrectableFS}
P_e<k\left(\frac{k}{q^m}\right)^{d_I}=k\left(\frac{k}{q^m}\right)^{s(\degConstraint-k+1)-\insertions(h-s+1)}
\end{equation}
under the assumption that the coefficients of the polynomials $B_0^{(1)}(x),\dots,B_0^{(d_I)}(x)$ are independent and uniformly distributed over $\Fqm$.
\end{lemma}

\begin{IEEEproof}
Evaluating $B_0^{(\ell)}(x),\ell\in\intervallincl{1}{d_I}$ at the distinct elements $\pe,\pe^{[1]},\dots,\pe^{[k-1]}$ gives a codeword of a $(k,s)$ Reed-Solomon code $\mathcal{C}_{RS}$. 
The probability to get a unique solution is then equal to the probability to get a weight $k$ codeword.
Since similar to \cite{BartzWachterInterleavedSubspace2014} and \cite{WachterzehZeh-InterpolationInterleavedGabidulin} we assume that the coefficients of $B_0^{(\ell)}(x),\ell\in\intervallincl{1}{d_I}$ are independent and uniformly distributed over $\Fqm$.
Hence we get a uniform distribution over the code book of $\mathcal{C}_{RS}$.
Using the approximation from \cite[Equation~1]{CheungWeightDist1989} the probability $P_\text{s}$ to get a codeword of full weight $k$ is 
\begin{align*}
 P_\text{s}
		  &\approx\frac{\text{no. of vectors of weight }k \text{ in }\Fqm^k}{\text{total no. of vectors in }\Fqm^k}=\left(1-\frac{1}{q^m}\right)^k\!.
\end{align*}
The probability that one $B_i^{(\ell)}(\pe^{[j]})$ in $\vec{b}_{0,i}$ is zero is at most $1-P_\text{s}$.
The probability that one $\vec{b}_{0,i}=\vec{0}, i\in\intervallincl{0}{k-1}$ is upper bounded by
\vspace*{-5pt}
\begin{equation*}
 \text{Pr}\left[\vec{b}_{0,i}\!=\!\vec{0}\right]<(1-P_\text{s})^{d_I}=\left(1\!-\!\left(1\!-\!\frac{1}{q^m}\right)^k\right)^{d_I}\!<\left(\frac{k}{q^m}\right)^{d_I}\!\!.
\end{equation*}
The probability that at least one $\vec{b}_{0,i}=\vec{0}$ for $i=0,\dots,k-1$ is thus upper bounded by
\begin{equation*}
 P_e<\text{Pr}\left[\bigcup_{i=0}^{k-1}\vec{b}_{0,i}=\vec{0}\right] \leq \sum_{i=0}^{k-1} \left(\frac{k}{q^m}\right)^{d_I} \!= k\left(\frac{k}{q^m}\right)^{d_I}\!\!.
\end{equation*}
\end{IEEEproof}

We restrict $d_I$ to be larger than a threshold $\minDim$, i.e. $\mu\leq d_I$, and get
\begin{equation}\label{eq:dIConditionFS}
			\degConstraint s\geq \insertions(h-s+1)+s(k-1)+\minDim.
\end{equation}
To ensure that $f(x)$ is a root of $P(x)$ in \eqref{eq:rootFindingPolyFS} the degree $\degConstraint$ must satisfy \eqref{eq:decConditionFS}, i.e.
	$\degConstraint\leq(\nTransmit-\deletions)(h-s+1)$.
By combining \eqref{eq:decConditionFS} and \eqref{eq:dIConditionFS} we get
\begin{equation}\label{eq:decRadiusFSunique}
 \insertions+s\deletions \leq \frac{s(\nTransmit(h-s+1)-(k-1))-\minDim}{h-s+1}.
\end{equation}
Under the assumption that the coefficients of the polynomials $B_0^{(\ell)}(x),\ell\in\intervallincl{1}{d_I}$ are independent and uniformly distributed over $\Fqm$ we can use Lemma~\ref{lem:failProb} to upper bound the failure probability.
If $\insertions$ and $\deletions$ fulfill \eqref{eq:decRadiusFSunique} we can find a unique solution $f(x)$ satisfying \eqref{eq:rootFindingPolyFS} with probability at least
\begin{equation*}
   1-k\left(\frac{k}{q^m}\right)^\minDim.
\end{equation*}
Using $R\approx\frac{(k-1)m}{\nTransmit(\nTransmit+hm)}$ the normalized decoding radius $\tau_u=\frac{\insertions+s\deletions}{\nTransmit}$ of the probabilistic unique decoding approach is
\begin{equation*}
 \tau_u\leq s\left(1-\frac{\nTransmit+hm}{m(h-s+1)}R\right)-\frac{\minDim}{(h-s+1)\nTransmit}.
\end{equation*}
In a setup where $\nTransmit h\approx m$ we have $\tau_u\leq s\big(1-\frac{1/h+h}{h-s+1}R\big)- \minDim/((h-s+1)\nTransmit)$. 

To adjust the decoding radius at the receiver we express the degree constraint $\degConstraint$ in terms of $\mu$.
Combining \eqref{eq:decConditionFS} and \eqref{eq:dIConditionFS} we get $\nReceive(h\!-\!s\!+\!1)\!+\!s(k\!-\!1)\!+\!\minDim\leq(s\!+\!1)(\nTransmit\!-\!\deletions)(h\!-\!s\!+\!1).$
From \eqref{eq:decConditionFS} we get $\degConstraint(s+1)\leq (s+1)(\nTransmit-\deletions)(h-s+1)$ and choose
\begin{align*}
	\degConstraint&=\Bigg\lceil\frac{\nReceive(h-s+1)+s(k-1)+\minDim}{s+1}\Bigg\rceil.
\end{align*}
The computational complexity of the unique decoder is dominated by the interpolation step, which can be solved requiring at most $\OCompl{s^2\nReceive \degConstraint(h-s+1)}<\OCompl{s^2\nReceive^2}$ operations in $\Fqm$ using the efficient algorithm in \cite{BartzWachterInterleavedSubspace2014}.
The pseudo code for the probabilistic unique decoder is given in Algorithm~\ref{alg:decodeFS}.

\begin{algorithm}
	\caption{UniqueDecodeFS$(\vec{x}^{T}, \vec{y}^{(1)T}, \dots, \vec{y}^{(h)T})$}\label{alg:decodeFS}
	\DontPrintSemicolon
	\SetKwInOut{Input}{Input}\SetKwInOut{Output}{Output}
	\Input{A basis $(\vec{x}^T, \vec{y}^{(1)T}, \dots, \vec{y}^{(h)T})$ for the $\nReceive$-dimensional received subspace}
	\Output{A polynomial $f(x)\in\Linpolyring:\deg(f(x))<k$ \\or ``decoding failure''}
	Set up $\Mat{T}\in\Fqm^{\nReceive(h-s+1)\times (s+1)}$ to contain all interpolation tuples of Problem~\ref{prob:intProblemFS} as rows
	and denote by $\vec{t}_0,\dots,\vec{t}_s$ the columns of $\Mat{T}$	\\
	\textbf{Interpolation step:} \\
	$Q^{(1)}\!,\dots,\! Q^{(d_I)}\!\gets\!$ InterpolateBasis$(\vec{t}_0^{T},\vec{t}_1^{T},\dots,\vec{t}_s^{T})$ 
	\textbf{Root-finding step:}\\
	$Q^*=\{Q^{(\ell)}:\deg_q(Q^{(\ell)})<\degConstraint,\ell\in\intervallincl{1}{d_I}\}$\\
	Set up the root-finding matrix $\Mat{B}$ as in \eqref{eq:rootFindingFS} using all polynomials in $Q^*$ \\
	\If{$\vec{b_{0,i}}\neq\vec{0}, \forall i\in\intervallincl{0}{k-1}$}{
		Solve $\Mat{B}\cdot\vec{f}=\vec{q}_{0}$ and define $f(x)$ from $\vec{f}$\\
		\textbf{Output:} $f(x)$
	}
	\Else{
		\textbf{Output:} ``decoding failure''
	}
\end{algorithm}

\subsection{Performance Analysis}
We compare the performance of our proposed code construction with the code constructions by Kötter and Kschischang \cite{koetter_kschischang}, Mahdavifar-Vardy \cite{Mahdavifar2012Listdecoding} and Guruswami-Xing \cite{GuruswamiXing-ListDecodingRSAGGabidulinSubcodes_2012}.
For a fair comparison we select the code parameters such that each codeword contains the same number of symbols.
Figure~\ref{fig:radiusFS} shows that the code by Mahdavifar and Vardy only can correct errors for very small rates.
The construction by Guruswami and Xing achieves the best decoding radius for all rates but puts out a very large list with high probability.

The proposed code construction can correct insertions and deletions for all code rates and returns a \emph{unique} solution with high probability, which is a major benefit for practical applications.
\begin{figure}[ht!]
    \centering
    \input{fracFSh10s4.tikz}
    \caption{The normalized decoding radius $\tau_f\!=\!\frac{\insertions\!+\!s\deletions}{\nTransmit}$ vs. the rate $R$ for $h\!=\!10$.}\label{fig:radiusFS}
\end{figure}

\subsection{Simulation Results}
Consider a folded subspace code with parameters $\nTransmit\!=\!3$, $h\!=\!3$, $k\!=\!4$, $s\!=\!2$, $q\!=\!2$ and $m\!=\!h\nTransmit\!=\!9$.
For $\mu\!=\!1$ (maximum decoding radius) we simulated $5.2\cdot 10^6$ transmissions over an operator channel with $\deletions\!=\!0$ deletions and $\insertions\!=\!2$ insertions and observed a fraction of $7.80\cdot10^{-3}$ decoding errors (upper bound $3.13\!\cdot\!10^{-2}$).
For $\mu\!=\!2$ the code corrects $\deletions\!=\!0$ deletions and $\insertions\!=\!2$ insertions.
We simulated $5.5\cdot10^6$ transmissions over an operator channel with parameters $\deletions\!=\!0,\insertions\!=\!2$ for $\mu\!=\!2$ and observed a fraction of $1.97\cdot10^{-5}$ decoding errors (upper bound $2.44\!\cdot\!10^{-4}$).
For $\mu\!=\!3$ the code can correct $\deletions\!=\!0$ deletions and $\insertions\!=\!1$ insertions.
After simulating $1.62\cdot 10^{7}$ transmissions for $\insertions=1$ we observed no decoding failure so far (upper bound $1.91\!\cdot\!10^{-6}$).
The simulation is still in progress.
The simulation results show that the assumptions in Lemma~\ref{lem:failProb} are reasonable. 

The decoder in \cite{Mahdavifar2012Listdecoding} can not correct any insertions and deletions for these code parameters, if the same number of symbols is transmitted.
The code from \cite{GuruswamiWang13Explicit} can correct the same number of insertions and deletions for the given parameters but will output a large list with high probability instead of a unique solution.

\section{Conclusion}\label{sec:conclusion}
A new family of folded subspace codes for error correction in noncoherent network coding scenarios was presented.
The codes are more resilient against injections of malicious packets (insertions).
An efficient interpolation-based decoding algorithm was presented that can be used as a list decoder or as a probabilistic unique decoder.
The decoder corrects insertions and deletions for any code rate.
We showed that the probabilistic unique decoder outputs a unique solution with high probability and gave an upper bound on the average list size and on the decoding failure probability.
The decoding radius of the unique decoder can be adjusted to control the decoding radius vs. failure probability tradeoff.

\section*{Acknowledgment} 
The authors would like to thank Gerhard Kramer and Joschi Brauchle for fruitful discussions and helpful comments.
H. Bartz was supported by the German Ministry of Education and Research in the framework of an Alexander von Humboldt-Professorship. 
V. Sidorenko is on leave from the Institute for Information Transmission Problems, Russian Academy of Science.


\bibliographystyle{IEEEtran}
\bibliography{references}

\end{document}

%% file: macros.tex
\definecolor{TUMdgray}{RGB}{088,088,090}
\definecolor{TUMmgray}{RGB}{156,157,159}
\definecolor{TUMlgray}{RGB}{215,217,218}
\definecolor{TUMyellow}{RGB}{255,180,000}
\definecolor{TUMorange}{RGB}{255,128,000}
\definecolor{TUMblue}{RGB}{000,101,189}
\definecolor{TUMgreen}{RGB}{0,124,48}
\definecolor{TUMred}{RGB}{196,7,27}
\definecolor{TUMgold}{RGB}{255,200,0}

\newtheorem{theorem}{Theorem}
\newtheorem{definition}{Definition}
\newtheorem{lemma}{Lemma}

\newtheorem{problem}{Problem}

\newcommand{\Fqm}{\ensuremath{\mathbb F_{q^m}}}

\newcommand{\Fq}{\ensuremath{\mathbb F_{q}}}

\newcommand{\intervallincl}[2]{\ensuremath{[#1,#2]}}
\newcommand{\intervallexcl}[2]{\ensuremath{[#1,#2-1]}}

\newcommand{\set}[1]{\ensuremath{#1}}

\newcommand{\Linpolyring}{\mathbb{L}_{q^m}\![x]}

\newcommand{\OCompl}[1]{\ensuremath{\mathcal{O}({#1})}}

\DeclareMathOperator{\defi}{def}
\newcommand{\defeq}{\overset{\defi}{=}}

\renewcommand{\bar}{\overline}

\DeclareMathOperator{\rk}{rk}

\renewcommand{\vec}[1]{\ensuremath{\mathbf{#1}}}
\newcommand{\Mat}[1]{\ensuremath{\mathbf{#1}}}
\newcommand{\vecelements}[1]{\ensuremath{(#1_0 \ #1_1 \ \dots \ #1_{n-1})}}

\newcommand{\Mooremat}[2]{\Mat{M}_{#1}( #2 )}

\newcommand{\qVan}{\emph{Moore} }

\newcommand{\MoormatExplicit}[3]{
\begin{pmatrix}
#1_{0} & #1_{1} & \dots& #1_{#3-1}\\
#1_{0}^{[1]} & #1_{1}^{[1]} & \dots& #1_{#3-1}^{[1]}\\[-4pt]
\vdots &\vdots&\ddots& \vdots\\[-2pt]
#1_{0}^{[#2-1]} & #1_{1}^{[#2-1]} & \dots& #1_{#3-1}^{[#2-1]}\\
\end{pmatrix}}

\newcommand{\FSub}[1]{\ensuremath{\mathrm{F}\mathrm{Sub}[#1]}}

\newcommand{\Subspacedist}[1]{d_s(#1)}

\newcommand{\myspace}[1]{\mathcal{#1}}
\newcommand{\Rowspace}[1]{\left\langle#1\right\rangle_{\!q}}

\newcommand{\Grassm}[1]{\myspace{G}_q(#1)}

\newcommand{\ProjspaceAny}[1]{\myspace{P}_q(#1)}

\newcommand{\volSubBall}[1]{\ensuremath{V_S(#1)}}
\newcommand{\numSubShell}[1]{\ensuremath{N_S(#1)}}

\newcommand{\quadbinom}[2]{\ensuremath{
{#1
\brack
#2}
}}

\newcommand{\nTransmit}{\ensuremath{n_t}}
\newcommand{\nReceive}{\ensuremath{n_r}}
\newcommand{\insertions}{\ensuremath{\gamma}}
\newcommand{\deletions}{\ensuremath{\delta}}
\newcommand{\intPoly}{\ensuremath{Q\left(x,y_1,\dots,y_s\right)}}

\newcommand{\pe}{\ensuremath{\alpha}}

\newcommand{\minDim}{\ensuremath{\mu}}
\newcommand{\degConstraint}{\ensuremath{D}}

%% file: fracFSh10s4.tikz
%
%
%
\begin{tikzpicture}

\begin{axis}[%
scale only axis,
xmin=0,
xmax=1,
xlabel={Code Rate $R$},
xmajorgrids,
ymin=0,
ymax=10,
ylabel={Normalized decoding radius $\tau_f$},
ymajorgrids,
legend style={legend cell align=left,align=left,draw=white!15!black}
]
\addplot [color=\KK,dotted, line width=1.2pt, mark=x]
  table[row sep=crcr]{%
0	1\\
0.1	0.9\\
0.2	0.8\\
0.3	0.7\\
0.4	0.6\\
0.5	0.5\\
0.6	0.4\\
0.7	0.3\\
0.8	0.2\\
0.9	0.1\\
1	0\\
};
\addlegendentry{Kötter-Kschischang};

\addplot [color=\MV,dashed, line width=1.1pt]
  table[row sep=crcr]{%
0	10\\
0.1	0\\
0.2	-10\\
0.3	-20\\
0.4	-30\\
0.5	-40\\
0.6	-50\\
0.7	-60\\
0.8	-70\\
0.9	-80\\
1	-90\\
};
\addlegendentry{Mahdavifar-Vardy};

\addplot [color=\GX,dashdotted, line width=1.1pt,mark=diamond]
  table[row sep=crcr]{%
0	10\\
0.1	9\\
0.2	8\\
0.3	7\\
0.4	6\\
0.5	5\\
0.6	4\\
0.7	3\\
0.8	2\\
0.9	1\\
1	0\\
};
\addlegendentry{Guruswami-Xing};

\addplot [color=\FS,solid,line width=1.1pt]
  table[row sep=crcr]{%
0	10\\
0.1	5.33333333333333\\
0.2	3.6\\
0.3	2.5\\
0.4	1.71428571428571\\
0.5	1.14285714285714\\
0.6	0.75\\
0.7	0.444444444444444\\
0.8	0.222222222222222\\
0.9	0.1\\
1	0\\
};
\addlegendentry{Folded Subspace Codes};

\end{axis}
\end{tikzpicture}%